\documentclass[12pt]{article} 
\usepackage[tablesfirst,notablist,nomarkers]{endfloat} 

\usepackage{ol2}
\usepackage[draft]{hyperref}
\usepackage{amsmath}

\begin{document}


\title{Surface plasmon-mediated far-field emission of laser dye solutions}


\author{Carola Geiger and Jochen Fick$^{*}$ }

\address{Institut N\'{e}el, CNRS and Universit\'{e} Joseph Fourier, BP 166, 38042 Grenoble, France \\
$^*$Corresponding author: jochen.fick@grenoble.cnrs.fr}

\begin{abstract}Angle-resolved reflection and emission spectra of metal gratings consisting of sub-wavelength grooves and immersed into rhodamine B and rhodamine 19 solutions are presented. The measured reflection and emission dispersion diagrams reveal the surface plasmon polaritons positions and strong plasmon mediated emission enhancement, respectively. The same grating could be easily re-used for the characterization of different dye-molecules.
\end{abstract}



Surface plasmon-based devices are of great interest for their potential applications in high sensitivity sensing \cite{AHL+08,HMJ+08}, improved light sources \cite{OSN+08} or in integrated optics devices \cite{Ozb06}. Amongst different plasmonic systems such as metal nano-particles or aperture arrays, metal gratings are very efficient for coupling radiative light to surface plasmon polaritons (SSP). As examples, the enhancement due to metal gratings was studied for organic emitters \cite{NCH+09,WMW+08}, dye molecules \cite{OHK04,GRV08}, J-aggregated cyanine dye layers \cite{BBS+06}, erbium doped silica glass \cite{KSG+03} or top-emitting organic light-emitting diodes \cite{WGB07}.

In these papers, the active molecules were integrated in thin films which were deposited onto a grating. The gratings are, in general, used only once. In this letter we report the investigation of a metal grating emerged into different dye solutions. This very flexible method allows to re-use gratings with different molecules or different dye concentrations. The main benefits of this approach are the possibility to study an enlarged spectral region by using several dyes, to optimize the dye concentration, or to make a quick check of the grating properties. The later point is interesting, as the elaboration of the sub-wavelength groove gratings represent a crucial and time consuming part of the work.


\begin{figure}[htb]
    \centerline{\includegraphics[width=8.3 cm] {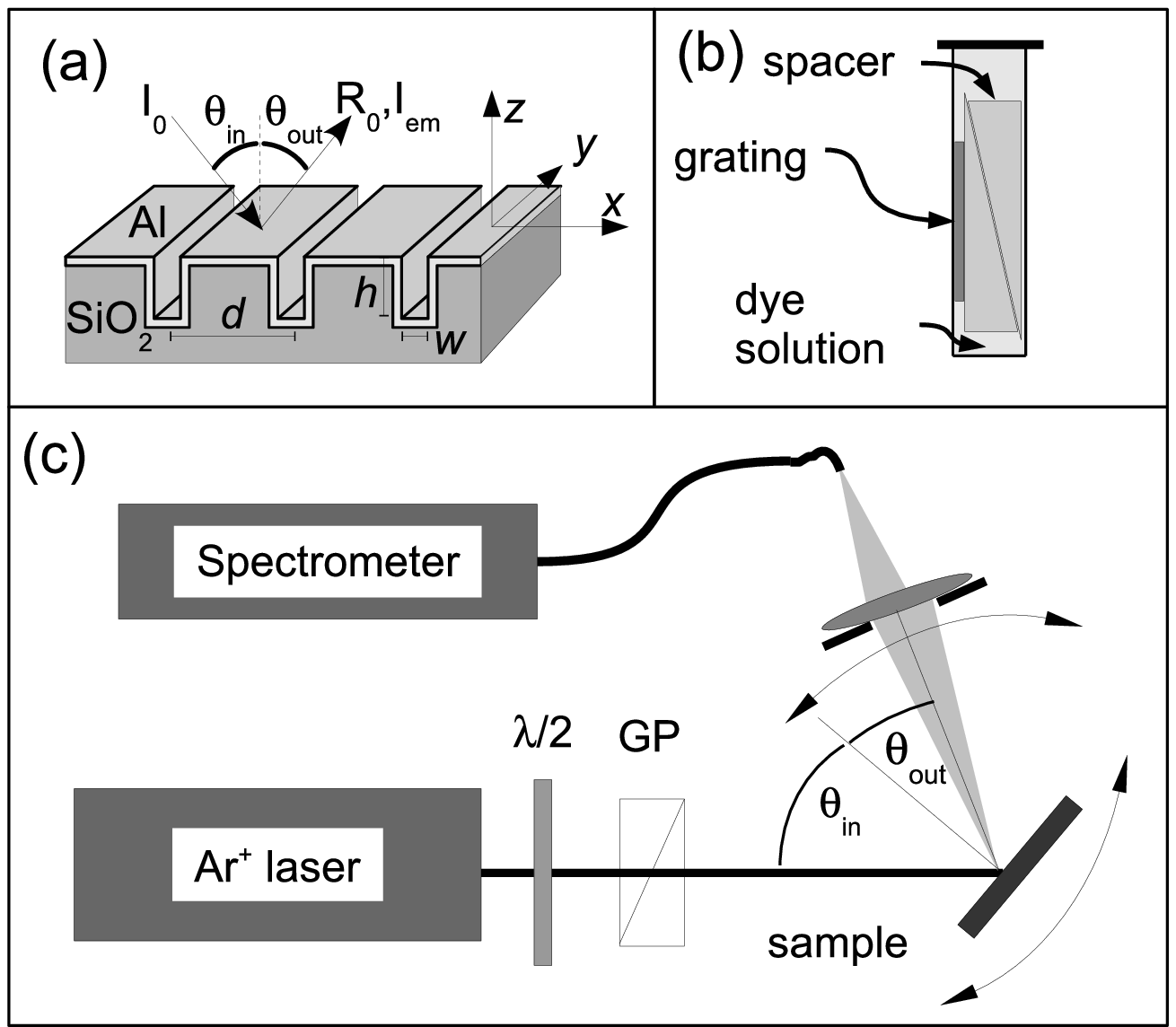}}
    \caption{(a) Schema of investigated sub-wavelength aluminum grating, (b) zoom on grating mount in the precision cell and (c) angle-resolved emission measurement set-up. \label{fig.exp}}
\end{figure}

Metal gratings were fabricated by standard electron beam lithography as described in \cite{BQB+03}. We used an aluminum grating with period $d=600$ nm, groove width $w=150$ nm and groove height $h\approx 600$ nm (Fig. \ref{fig.exp}.a). This class of gratings supports two different kinds of surface plasmon polariton (SPP) modes: surface modes (SM) at the grating surface and cavity modes (CM) inside the grating grooves.

Modal theory was applied for the calculation of the SPP resonance positions \cite{BQB+03}.  Our model is based on the development of the electromagnetic field inside the grating grooves into a set of guided or evanescent SPP modes. At resonance wavelengths $\lambda^{m}\approx 4h/(2m-1)$ these modes form Fabry-Perot type cavity modes CM$^{(m)}$. The SPP modes at the grating surface SM$^{(m)}$ are described by their dispersion relation : $k_{||}^{SM}=k_{SPP}(sin\theta^{(m)} +m\lambda/d)$ with $\theta$ the incident angle, $k_{||}^{SM}$ the SM$^{(m)}$ wavevector in-plane component, and $k_{SPP}=2 \pi / \lambda \sqrt{\epsilon_m/1+\epsilon_m)}$ the SPP wavewector at a plane metal surface. The applicability of this model to sub-wavelength metal gratings in the visible spectral region was proven \cite{ MFJ09}.

Angle-resolved reflection measurements were performed using the set-up described in \cite{MFJ09}. All measurements are made in TM (or p) polarization. They are normalized by the planar metal surface reflection, measured beside the grating. Dispersion diagrams are calculated from reflection spectra by constructing a color plot representing the reflected intensity as a function of the normalized in plane wave vector $k_{||}=k_0 (d/\pi) \sin \theta$ and the wave number $wn=1/\lambda$.

The experimental set-up used for angle-resolved emission measurements is shown on Fig. \ref{fig.exp}.c. A continuous wave Spectra Physics Ar$^+$ ion laser emitting at 514 or 488 nm wavelength is used as excitation source. The beam radius is $4$ mm with incident intensities of $1-1.5$ mW. The pump beam was TE polarized to allow efficient residual pump light suppression for TM emission. The sample is mounted in the center of a homemade $\theta/2\theta$ goniometer and its emission is coupled into a multimode optical fiber by a lens mounted on the goniometer branch. The spectra are recorded using a  HR4000 spectrometer from Ocean Optics. The used integration time was 300 ms and the spectra were averaged over three measurements. Emission spectra were recorded for output angles of up to $\theta_{out}=\pm75^\circ$ at an input angle of typically  $\theta_{in}=30^\circ$.

The samples were immersed into the dye solutions inside a rectangular precision cell from Hellma (Fig. \ref{fig.exp}.b). The gratings were pressed to the front side of the cell by  polystyrene-wedges. To avoid evaporation of the dye-solution, the cell was sealed using modeling clay. Before changing the dye solution the grating is rinsed with methanol. The influence of the cell was investigated by recording reflection measurements for gratings inside and without the precision cell. No significant degradation could be found. Only the reflected intensity decreased slightly and the angular range was limited to $75^\circ$ instead of $85^\circ$.

Rhodamine 19 (R19) and rhodamine B (RB) dissolved in methanol were used as laser dye solutions. The respective absorption and emission maxima are 520/ 545nm for R19 and 548/ 570nm for RB. These dyes can be easily pumped with an Ar$^+$ laser and their combined emission spectra allows to study the 525-600 nm wavelength range. The concentration was chosen to 0.133 g/L, so that the dye absorption is of the order of the losses due to plasmon resonance coupling.

\begin{figure}[htb]
      \centerline{\includegraphics[width=8.3 cm] {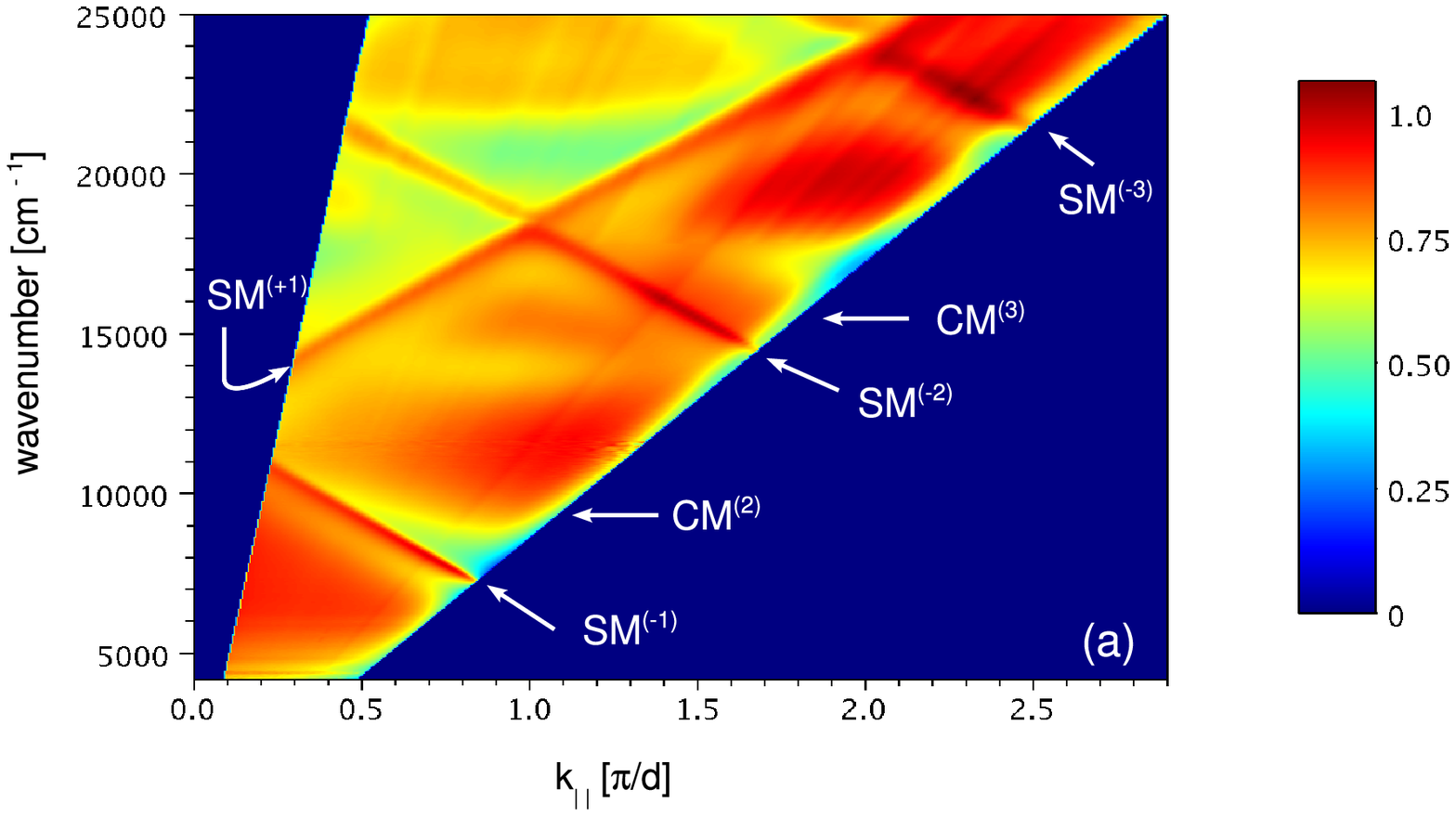}}
      \centerline{\includegraphics[width=8.3 cm] {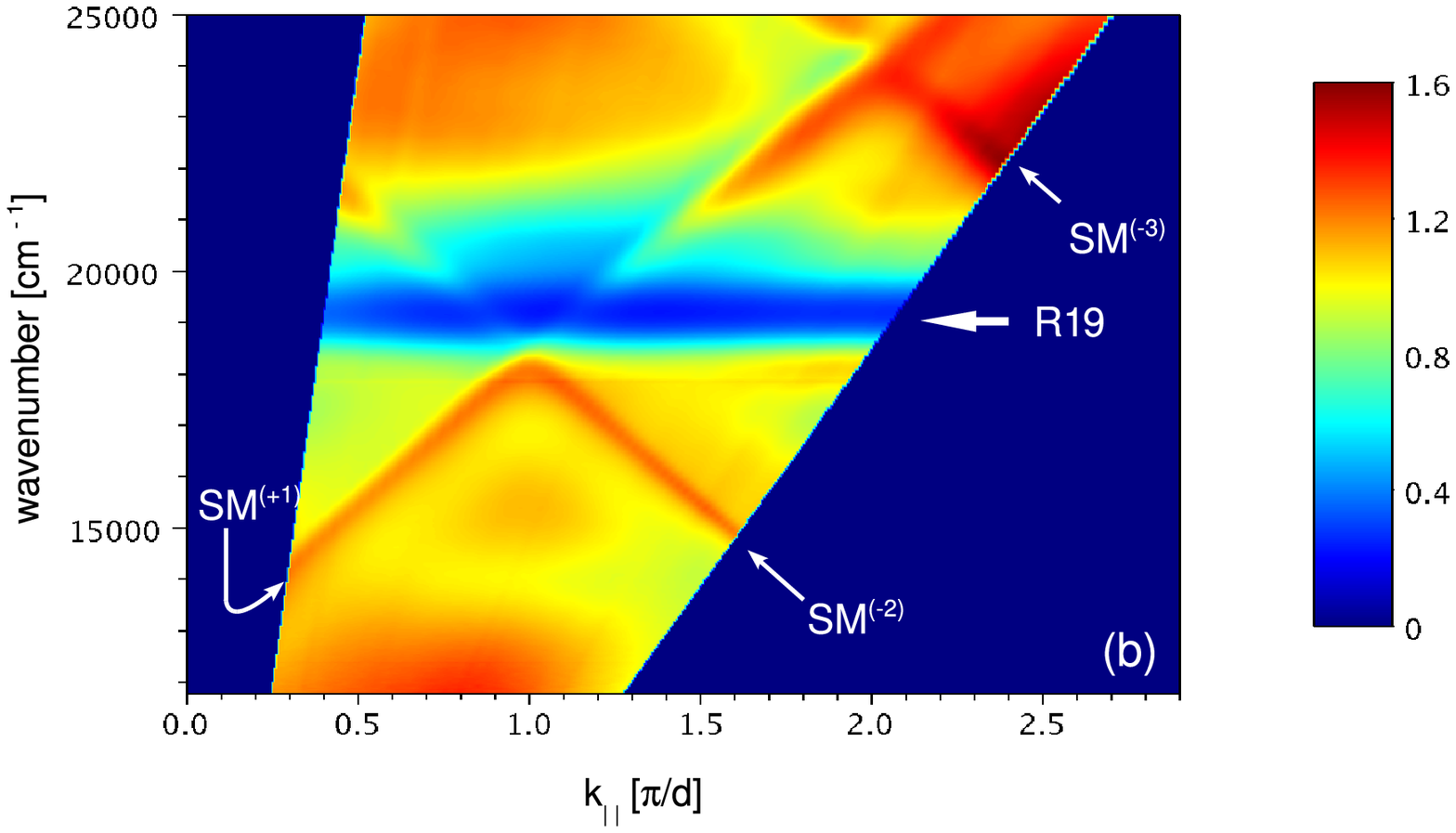}}
    \caption{Reflection dispersion diagram of the aluminum grating in methanol (a) and R19-methanol solution (b).\label{fig.disp}}
\end{figure}

The dispersion diagram of the reflection measurements of the grating immersed in methanol is shown in Fig. \ref{fig.disp}.a. The surface modes SM$^{(-1)}$ to SM$^{(-3)}$ and SM$^{(+1)}$ can be clearly distinguished. They are represented by a network of inclined straight lines. These dark lines correspond to local reflection minima due to intrinsic plasmon mode losses at the metal surface. The cavity modes CM$^{(2)}$ and  CM$^{(3)}$ which should appear around 9000 and 15500 cm$^{-1}$, respectively, could not be resolved. The fundamental mode CM$^{(1)}$, which is in general the most visible mode, is situated at $\approx3000$ cm$^{-1}$, below the experimentally accessible region.
 
The dispersion diagram of the grating immersed in the R19 methanol solution is shown on Fig. \ref{fig.disp}.b. Here, the reflection of the plane metal surface in pure methanol was used for normalization. The absorption of the laser dye around 19200 cm$^{-1}$ is clearly visible. As expected, the positions of the plasmon resonances are the same as for pure methanol. Only in the R19 absorption region the plasmon resonances become nearly invisible.

\small
\begin{figure}[htb]
\centerline{\includegraphics[width=8.3cm]{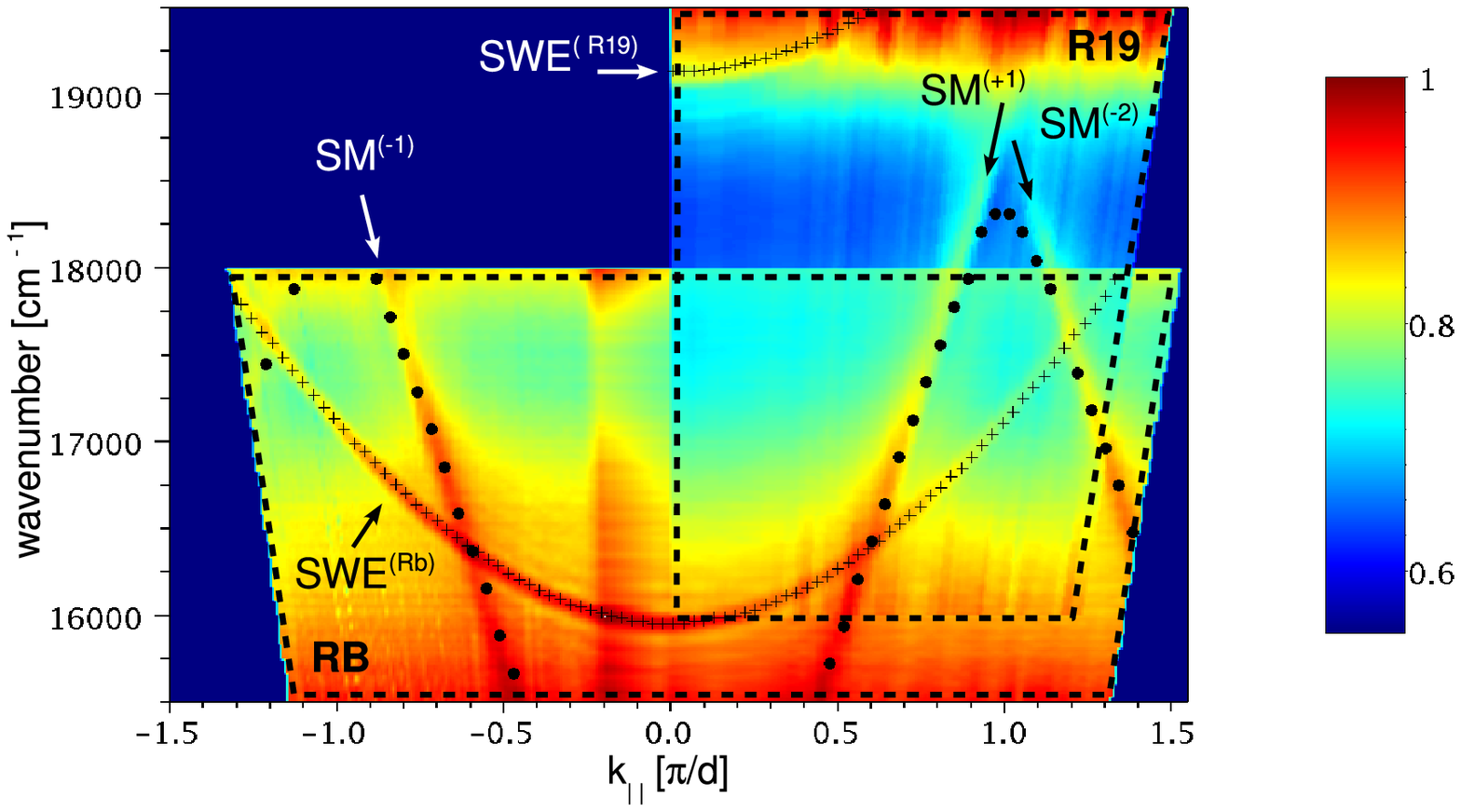}}
 \caption{Emission dispersion diagram obtained from two separate measurement series with RB and R19 solutions. ($\bullet$ : calculated position of the SM modes, + : calculated dispersion of second order standing wave enhancement.)\label{fig.em}}
\end{figure}
\normalsize

The emission dispersion diagram obtained from successive emission measurements of the grating immersed into R19 and Rb solutions is shown in Fig. \ref{fig.em}. The spectra were normalized by the emission, measured on the plane metal surface just beside the grating. In the overlap region the respective results were added and divided by two. For wavelengths outside the dye emission region, the measured signals vanish. Thus the normalized emission $\approx 1$, as can be seen for wavenumbers above 19000 and below 15750 cm$^{-1}$.

Two types of emission enhancement features with distinct dispersions can be observed: SPP enhanced emission (straight lines) and standing wave enhancement (SWE, parabolas). Outside of the enhancement features, the normalized emission is below 1. This effect is explained by emission concentration into the enhanced directions, but also by diffraction and absorption losses due to irregularities and poorer metal quality in the grating region.

The SWE is independent from SPP modes as it was observed even for a plane metal surface and for TE emission polarization. Closer investigation revealed a wavelength dependency on the emission and excitation angles which could be approximated by $\lambda_{SWE}\propto\cos{\theta_{out}^{d}}/\cos{\theta_{in}^{d}}$ ($\theta^{d}$ : angle in the dye-solution).

The emission angle dependency can be explained by vertical standing wave modes at the emission wavelength. The corresponding Fapry-Perot-type cavity was built by the metal surface and the precision cell inner surface. The high local intensity of the standing wave modes enables stimulated emission, which increases the overall emitted light intensity. The points in Fig. \ref{fig.em} were calculated for the second order standing waves modes with a dye solution thicknesses of 925 nm/ 770 nm for the RB and R19 solutions, respectively.

In a similar way, the incident angle dependency is related to standing wave modes at the pump wavelength. To explain this effect, the emission of a single dipole above a metal mirror has to be considered. The light intensity emitted into a specific angle is composed by the interference between the direct "up-ward" emission and the "down-ward" emission reflected by the mirror  \cite{Dre70}. This interference leads to an angular emission distribution, which depends on the dipole/ mirror distance $d_{dm}$. In a homogeneously pump medium the integration over the emission of dipoles with continuous $d_{dm}$ obviates the observation of any interference figure. In the present case, the standing wave modes results in non-homogeneous pumping, which can lead to incident angle dependent preferential emission angles. We confirmed this interpretation by simulations based  on our modal model of the metal grating and using the Weyl development of the electric dipole radiation \cite{ERS99}.

SPP - laser dye coupling is revealed by the enhanced emission at the position of the SM$^{(\pm1)}$ and SM$^{(-2)}$ modes. The crosses shown in the figure correspond to calculated values of the SM mode position using the grating parameters determined in \cite{MFJ09} ($d= 600$nm, $w=150$nm, and $h=630$nm). This feature was only visible for TM polarized emission, evidencing that the emission enhancement is due to plasmon coupling and not to coupling into the grating diffraction modes. The position, contrast, and width of the plasmon enhanced emission lines of the two dye solutions agree very well.

In our experiment, the distance between the grating surface and the precision cell, i.e. the active media thickness, is only roughly controlled. This fact was confirmed by the estimated thicknesses from the SWE dispersion. However the SPP enhancement is independent from these thickness variations. In fact, the emission enhancement is due to near field coupling of the dipoles excitation to surface plasmon modes, followed by grating mediated coupling to the radiative far field. Only dye molecules close to the metal grating ($< 100$ nm) are concerned by this very  efficient near field coupling. More distant molecules can couple to the plasmon modes only by grating mediated far-field coupling. This latter effect represent, however, a supplementary loss of electromagnetic energy. Thus, the observed emission enhancement concerns only a very small part of dye-molecules. For this reason the measured relative emission enhancement is lower than for similar dye molecules but included in very thin films (20 - 50 nm) on metallic gratings  \cite{OHK04}, or planar films \cite{LSM+08}.

In conclusion, emission enhancement of dye molecules due to near-field plasmon coupling on sub-wavelength metal gratings was investigated. Reflection and emission dispersion diagrams with high resolution and good contrast were measured. The viability of our solution approach was demonstrated by using the same grating for the successive study of different dye-molecules.

\bibliographystyle{unsrt}

\end{document}